# Clinical Prediction Models to Predict the Risk of Multiple Binary Outcomes: a comparison of approaches

Glen P. Martin [1]; Matthew Sperrin [1]; Kym I.E. Snell [2]; Iain Buchan [3]; Richard D. Riley [2]

1. Division of Informatics, Imaging and Data Science, Faculty of Biology, Medicine and Health, University of Manchester, Manchester Academic Health Science Centre, Manchester, UK

2. Centre for Prognosis Research, School of Primary, Community and Social Care, Keele University, Staffordshire, UK

3. Institute of Population Health Sciences, Faculty of Health and Life Sciences, University of Liverpool, Liverpool, UK

**Running Title:** Approaches to clinical prediction with multiple binary outcomes

**Competing Interests:** None

**Corresponding Author**
Dr Glen Philip Martin
Lecturer in Health Data Science
Division of Informatics, Imaging and Data Science, Faculty of Biology, Medicine and Health, University of Manchester,
Vaughan House, Manchester, M13 9GB, United Kingdom
Email: glen.martin@manchester.ac.uk

**Abstract**

Clinical prediction models (CPMs) are used to predict clinically relevant outcomes or events. Typically, prognostic CPMs are derived to predict the risk of a single future outcome. However, with rising emphasis on the prediction of multi-morbidity, there is growing need for CPMs to simultaneously predict risks for each of multiple future outcomes. A common approach to multi-outcome risk prediction is to derive a CPM for each outcome separately, then multiply the predicted risks. This approach is only valid if the outcomes are conditionally independent given the covariates, and it fails to exploit the potential relationships between the outcomes. This paper outlines several approaches that could be used to develop prognostic CPMs for multiple outcomes. We consider four methods, ranging in complexity and assumed conditional independence assumptions: namely, probabilistic classifier chain, multinomial logistic regression, multivariate logistic regression, and a Bayesian probit model. These are compared with methods that rely on conditional independence: separate univariate CPMs and stacked regression. Employing a simulation study and real-world example via the MIMIC-III database, we illustrate that CPMs for joint risk prediction of multiple outcomes should only be derived using methods that model the residual correlation between outcomes. In such a situation, our results suggest that probabilistic classification chains, multinomial logistic regression or the Bayesian probit model are all appropriate choices. We call into question the development of CPMs for each outcome in isolation when multiple correlated or structurally related outcomes are of interest and recommend more holistic risk prediction.

# 1   Introduction

Prognostic clinical prediction models (CPMs) aim to predict the future occurrence of clinically relevant outcomes for an individual, given information known about them at the time of prediction [1–3]. CPMs are predominately derived in a multivariable regression framework (e.g. logistic regression for binary outcomes), which combine estimated associations between multiple predictors (risk or prognostic factors) and an outcome of interest.

CPMs can support clinical decision-making, but a single CPM is usually developed to predict only a single outcome and so seldom reflects actual healthcare pathways, since patients often develop multiple conditions and outcomes over time, leading to multi-morbidity. In situations where one wishes to predict the joint risk of multiple outcomes (e.g. in predicting future multi-morbidity, where interest is in the risk of a patient developing different combinations of conditions), such univariate CPMs are undesirable because they fail to reflect real-world health care and do not model dependency between clinical outcomes.

For example, suppose we are interested in predicting the risk of two different binary outcomes occurring, where the occurrence of one outcome (e.g. onset of type-2 diabetes) does not prevent – but might be associated with – the occurrence of the other outcome (e.g. onset of coronary heart disease). The standard approach to prediction modelling would be to derive a CPM for each of the two outcomes separately. If the joint risk of developing multiple outcomes (e.g. risk of developing both type-2 diabetes and coronary heart disease) is of interest, then the typical way to estimate this from univariate models is by multiplying their predicted risks. However, this approach only leads to reliable estimates of the joint risk if the outcomes are conditionally independent given the covariates; additionally, it fails to exploit the potential relationships between the outcomes, which could improve inference and joint risk prediction [4,5]. A common alternative is to create a composite outcome (defined as occurrence of either or both outcomes); for example, the definition of major adverse cardiovascular events [6]. While this approach simplifies the modelling, it is problematic to estimate marginal risks of each outcome (or different combinations of the outcomes co-occurring), leads to a loss of important information and has potential replication issues in situations where there is no universally agreed definition of the composite measure (e.g. [6]).

Therefore, as healthcare strives to move from being reactive (treat/manage established disease) to proactive (prevention and early detection), CPMs need to operate in a more holistic manner, especially with an increasing global priority for multi-morbidity prediction and management [7,8]. To this end, regression approaches that allow multiple outcomes to be modelled simultaneously have a long history in the statistical literature [4,5,9,10]. In the predictive modelling literature, multi-state models have been applied to predict the risk of patients moving between different states (of diseases/conditions/pathways) pertaining to a combination of different outcomes through time and accounting for competing risks [11,12]. Similarly, problems in text classification and annotation have spawned machine learning techniques based on multi-label classification/learning [13,14]. These include binary relevance [15], ensemble of classifier chains [16], multi-label decision trees [13] and multi-label neural networks [17].

Nonetheless, multiple outcome prediction methods are rarely utilised within the predictive modelling field [18], and the effects of ignoring dependency between outcomes on predictive performance has received little attention. In this study we propose a variety of approaches for developing prognostic CPMs for multiple binary outcomes and compare their performance through a simulation study and real-world example. We consider estimation of both marginal and joint probabilities of outcomes, and compare each method's ability to estimate these under different scenarios.

The remainder of the paper is structured as follows: in section 2 we outline notation and present current univariate approaches to developing CPMs (i.e. those that rely on conditional independence); we provide an overview of several methods to develop prognostic CPMs for multiple binary outcomes in section 3; in section 4 we describe the design and results of a simulation study comparing the methods, while in section 5 we apply the methods to a real-world critical care example; finally, in section 6 we discuss our findings and present directions for future work.

## 2 Prediction Approaches under Conditional Independence
### 2.1 Notation and Preliminaries

Throughout, we assume that the modeller has access to individual participant data (IPD) on a population of interest. The IPD includes $N$ independent observations of $P$ predictor variables, arranged in an $N \times P$ matrix $\boldsymbol{X} = (\boldsymbol{X}_1, \dots, \boldsymbol{X}_P)$, with the $(i,p)^{th}$ element of $\boldsymbol{X}$ denoted as $x_{i,p}$. Additionally, each observation within the IPD has a set of $K$ unique (but potentially related) binary outcomes, which we denote as $Y_{i1}, \dots, Y_{iK}$, where $Y_{ij} = 1$ if observation $i$ had the $j$-th outcome event, with $Y_{ij} = 0$ otherwise. We assume that occurrence of one outcome does not preclude the occurrence of any of the others (e.g. a setting of multi-morbidity prediction; thus excluding death). For ease of exposition we describe each of the methods in the case where $K = 2$, but extensions to $K > 2$ follow naturally.

Unless otherwise stated, each of the methods described below (in sections 2 and 3) are fitted using maximum likelihood estimation (MLE), and we assume that a suitable CPM development strategy is employed [1,2,19], which may include adjustment for overfitting (e.g. penalised MLE). As a general point, overfitting is an important aspect of CPM development that should be accounted for, but this is not the focus of the current paper. For example, many of the methods described below are fit using unpenalised MLE, but methods such as LASSO or fitting through Bayesian inference with penalising priors might be useful in some situations to minimise overfitting [2,20,21]. While we do not consider this here, we postulate that Bayesian inference might be particularly useful when one wishes to incorporate prior knowledge into the parameter estimation (e.g. if there are existing CPMs for the individual outcomes [22,23]).

### 2.2 Univariate CPMs

The classic approach is to develop a univariate CPM for each outcome separately, for example using logistic regression. Specifically, we have that

$$P(Y_{ij} = 1|X_i) = \left[1 + \exp\left(-(\beta_{0,j} + X_i\beta_j)\right)\right]^{-1}, \quad (1)$$

for $j = 1, \ldots, K$, where $\beta_{0,j}$ is the intercept and $\boldsymbol{\beta_j} = (\beta_{1,j}, \ldots, \beta_{P,j})$ is a vector of coefficients, which represent the adjusted prognostic effects of the predictors on the $j$-th outcome. Here, $\beta_{0,j} + X_i\beta_j$ is referred to as the linear predictor. In this study, we fit equation (1) using MLE, noting that, in practice, overfitting should be accounted for in some situations.

At the time of making a prediction for a new individual with covariates $X_i^*$, the marginal predicted risk for the $j$-th outcome is $P(Y_{ij} = 1|X_i^*)$, while the joint probability $P(Y_{i1} = 1, \ldots, Y_{iK} = 1|X_i^*)$ is $\prod_{j=1}^{K} P(Y_{ij} = 1|X_i^*)$, meaning the approach relies on conditional independence of the outcomes to be valid.

### 2.3 Stacked Regression

One way to extend the univariate approach is based on the stacked regression literature [23,24], which allows the model for one of the outcomes to exploit the information (i.e. predictor-outcome associations) contained within the other outcome models, thereby improving marginal risk prediction. This approach can also be used in a setting where there are existing univariate CPMs available in the literature [23,25,26]. Stacked regression is a two-stage approach to model fitting, whereby individual CPMs are fit to each outcome independently using the IPD (or obtained from the literature), the linear predictors of which are then used in a second stage as covariates in a stacked regression model for each outcome.

Specifically, with $K = 2$, in the first stage we have that $\hat{f}_1(X) = \beta_{01} + X\boldsymbol{\beta}_1$ and $\hat{f}_2(X) = \beta_{02} + X\boldsymbol{\beta}_2$ (each estimated using unpenalised maximum likelihood estimation of equation (1), or obtained from the literature [23,25,26]). Then, in the second stage, we fit the following models in the IPD:

$$P(Y_{i1} = 1|X_i) = \left[1 + \exp\left(-\left(\hat{\eta}_{0,1} + \hat{\eta}_{1,1}\hat{f}_1(X_i) + \hat{\eta}_{2,1}\hat{f}_2(X_i) + \sum_{p=1}^{P} \hat{\delta}_{p,1} x_{i,p}\right)\right)\right]^{-1}, \quad (2)$$

and

$$P(Y_{i2} = 1|X_i) = \left[1 + \exp\left(-\left(\hat{\eta}_{0,2} + \hat{\eta}_{1,2}\hat{f}_1(X_i) + \hat{\eta}_{2,2}\hat{f}_2(X_i) + \sum_{p=1}^{P} \hat{\delta}_{p,2} x_{i,p}\right)\right)\right]^{-1}. \quad (3)$$

The unknown parameters $\hat{\eta}_{1,1}, \hat{\eta}_{2,1}, \hat{\eta}_{1,2}, \hat{\eta}_{2,2}, \hat{\delta}_{p,1}$ and $\hat{\delta}_{p,2}$, in equations (2) and (3) are estimated by maximising a lasso penalised likelihood [27], as previously described [23]. The final summations in equations (2) and (3) allow the individual predictor effects to differ dependent on the inclusion of $\hat{f}_1(X_i)$ and $\hat{f}_2(X_i)$ in each model.

At the time of prediction, the predicted marginal risks for the $j$-th outcome for a new individual with covariates $X_i^*$, are obtained by calculating $\hat{f}_1(X_i^*)$ and $\hat{f}_2(X_i^*)$, using the models obtained in the first stage, and then inserting these into the appropriate stacked regression model from the second stage (i.e. equation (2) or (3)). We consider this approach since it is a previously proposed alternative to univariate CPMs, but note that the joint probabilities for all outcomes, $P(Y_{i1} = 1, \ldots, Y_{iK} = 1|X_i)$ is computed as $\prod_{j=1}^{K} P(Y_{ij} = 1|X_i)$, meaning the approach still relies on conditional independence of the outcomes to be valid.

## 3 Prediction Approaches Accounting for Conditional Dependence

In this section, we describe four approaches that can readily be applied to real-world data that relax the conditional independence assumption to enable joint outcome risk estimation. In all the approaches, we allow for differences in the predictors for each of the outcomes since some elements of $\boldsymbol{\beta}_j$ could be estimated (or fixed) to be zero. To reiterate, our goal is to propose methodology that could develop a combinatorial network of CPMs, which respects the interplay between predictor effects and risks of multiple outcomes; this is key for accurate multi-morbidity risk prediction.

### 3.1 Probabilistic Classifier Chains

If the risks of multiple outcomes are related to each other (after conditioning on the covariates), then one approach to modelling dependence is to condition sequentially on each outcome. Consider the (randomly indexed) sequence of outcomes $Y_{i1}, \ldots, Y_{iK}$, then one can relax the conditional independence assumption by conditioning on $Y_{i1}, \ldots, Y_{ij-1}$ when predicting $Y_{ij}$ instead of only the covariates. Since the order of the indexing (of $Y_{i1}, \ldots, Y_{iK}$) will affect inference, an iterative approach is used, whereby all the permutations of the ordering of $Y_{i1}, \ldots, Y_{iK}$ are considered. As $K$ increases, so too does the number of 'permutations'; here, one could pick a random sample of permutations (similar to a permutation test), rather than fitting models on all $K!$ permutations.

Specifically, where $K = 2$, the first 'permutation' (denoted by a superscript (1)) is such that

$$P(Y_{i1} = 1|X_i) = \pi_{i1}^{(1)} = \left[1 + \exp\left(-\left(\beta_{0,1}^{(1)} + X_i\boldsymbol{\beta}_1^{(1)}\right)\right)\right]^{-1}$$

$$P(Y_{i2} = 1|X_i, Y_{i1}) = \pi_{i2}^{(1)} = \left[1 + \exp\left(-\left(\beta_{0,2}^{(1)} + X_i\boldsymbol{\beta}_2^{(1)} + \gamma_1^{(1)}Y_{i1}\right)\right)\right]^{-1},$$

while the second 'permutation' (denoted by a superscript (2)) is such that

$$P(Y_{i2} = 1|X_i) = \pi_{i2}^{(2)} = \left[1 + \exp\left(-\left(\beta_{0,2}^{(2)} + X_i\boldsymbol{\beta}_2^{(2)}\right)\right)\right]^{-1}$$

$$P(Y_{i1} = 1|X_i, Y_{i2}) = \pi_{i1}^{(2)} = \left[1 + \exp\left(-\left(\beta_{0,1}^{(2)} + X_i\boldsymbol{\beta}_1^{(2)} + \gamma_2^{(2)}Y_{i2}\right)\right)\right]^{-1}.$$

All models are fitted separately using maximum likelihood estimation (i.e. the models in the first 'permutation' are fitted independently of the models in the second 'permutation'). This approach is based on ensemble probabilistic classification chains from the multi-label classification literature [16,28].

Importantly, the conditioning on 'preceding' outcomes allows us to derive analytical expressions for the joint probabilities using Bayes' rule and by taking an average ensemble across the permutation models. For example, with $K = 2$ we have the following (omitting the conditions on $\boldsymbol{X}_i$ for notational brevity):

$$P(Y_{i1} = 1, Y_{i2} = 1) = \frac{1}{2}[P(Y_{i2} = 1|Y_{i1} = 1)P(Y_{i1} = 1) + P(Y_{i1} = 1|Y_{i2} = 1)P(Y_{i2} = 1)]$$
$$= \frac{1}{2}\left[\pi_{i2}^{(1)}\pi_{i1}^{(1)} + \pi_{i1}^{(2)}\pi_{i2}^{(2)}\right],$$

$$P(Y_{i1} = 1, Y_{i2} = 0) = \frac{1}{2}[P(Y_{i2} = 0|Y_{i1} = 1)P(Y_{i1} = 1) + P(Y_{i1} = 1|Y_{i2} = 0)P(Y_{i2} = 0)],$$

$$P(Y_{i1} = 0, Y_{i2} = 1) = \frac{1}{2}[P(Y_{i2} = 1|Y_{i1} = 0)P(Y_{i1} = 0) + P(Y_{i1} = 0|Y_{i2} = 1)P(Y_{i2} = 1)],$$

$$P(Y_{i1} = 0, Y_{i2} = 0) = \frac{1}{2}[P(Y_{i2} = 0|Y_{i1} = 0)P(Y_{i1} = 0) + P(Y_{i1} = 0|Y_{i2} = 0)P(Y_{i2} = 0)],$$

from which one can calculate the associated marginal probabilities of $P(Y_{i1} = 1|\boldsymbol{X}_i)$ and $P(Y_{i2} = 1|\boldsymbol{X}_i)$.

When predicting for a new individual (where clearly their outcomes are unknown), one would use the derived models to calculate the joint probabilities for all possible outcome combinations $Y_{ij} \in \{0,1\}$ for $j = 1, \dots, K$, which by definition sum to one. Computational tractability of this depends on small $K$.

### 3.2 Multinomial Logistic Regression

A second approach to modelling outcome dependence, which allows calculation of the predicted risk of different combinations of $Y_{ij} \in \{0,1\}$ for $j = 1, \dots, K$, is to use multinomial logistic regression, where the $2^K$ combinations are each treated as a nominal outcome category. For example, with two outcomes ($K = 2$), we fit the following models:

$$\log\left(\frac{P(Y_{i1} = 1, Y_{i2} = 1)}{P(Y_{i1} = 0, Y_{i2} = 0)}\right) = \beta_{0,1} + \boldsymbol{X}_i\boldsymbol{\beta}_1$$

$$\log\left(\frac{P(Y_{i1} = 1, Y_{i2} = 0)}{P(Y_{i1} = 0, Y_{i2} = 0)}\right) = \beta_{0,2} + \boldsymbol{X}_i\boldsymbol{\beta}_2$$

$$\log\left(\frac{P(Y_{i1} = 0, Y_{i2} = 1)}{P(Y_{i1} = 0, Y_{i2} = 0)}\right) = \beta_{0,3} + \boldsymbol{X}_i\boldsymbol{\beta}_3$$

These models are estimated using iterative procedures to find numerical optimisation of the parameters; practically, such models can be fit using R with the package nnet [29]. This approach becomes computationally intensive as the number of outcomes (and therefore outcome combinations) increases.

At the time of prediction for a new individual with covariates $\boldsymbol{X}_i^*$, we use the fact that the probabilities must sum to one, to allow us to explicitly obtain each joint probability. For the case with $K = 2$ we have

$$P(Y_{i1} = 1, Y_{i2} = 1 | X_i^*) = \frac{\exp(\beta_{0,1} + X_i^* \boldsymbol{\beta}_1)}{1 + \sum_{k=1}^{3} \exp(\beta_{0,k} + X_i^* \boldsymbol{\beta}_k)},$$

$$P(Y_{i1} = 1, Y_{i2} = 0 | X_i^*) = \frac{\exp(\beta_{0,2} + X_i^* \boldsymbol{\beta}_2)}{1 + \sum_{k=1}^{3} \exp(\beta_{0,k} + X_i^* \boldsymbol{\beta}_k)},$$

$$P(Y_{i1} = 0, Y_{i2} = 1 | X_i^*) = \frac{\exp(\beta_{0,3} + X_i^* \boldsymbol{\beta}_3)}{1 + \sum_{k=1}^{3} \exp(\beta_{0,k} + X_i^* \boldsymbol{\beta}_k)},$$

$$P(Y_{i1} = 0, Y_{i2} = 0 | X_i^*) = \frac{1}{1 + \sum_{k=1}^{3} \exp(\beta_{0,k} + X_i^* \boldsymbol{\beta}_k)}.$$

### 3.3 Multivariate Logistic Regression

Our third approach, which has previously been described in the context of modelling correlated binary outcomes, makes explicit use of a multivariate logistic distribution [30]. For ease of exposition, we again describe the case when $K = 2$ and refer readers to the literature for the more general case [30]. Explicitly, we use the bivariate logistic distribution proposed by Gumbel [31] to set

$$P(Y_{i1} = 1, Y_{i2} = 1 | X_i) = F_{i1} F_{i2} + \rho \sqrt{F_{i1} S_{i1} F_{i2} S_{i2}}$$

where $F_{ij} = P(Y_{ij} = 1 | X_i)$ for $j = 1,2$ (as defined in equation 1) and $S_{ij} = 1 - F_{ij}$. Here, $\rho$ estimates the residual correlation between the outcomes. Similarly, we have the following:

$$P(Y_{i1} = 1, Y_{i2} = 0 | X_i) = F_{i1} S_{i2} - \rho \sqrt{F_{i1} S_{i1} F_{i2} S_{i2}}$$

$$P(Y_{i1} = 0, Y_{i2} = 1 | X_i) = S_{i1} F_{i2} - \rho \sqrt{F_{i1} S_{i1} F_{i2} S_{i2}}$$

$$P(Y_{i1} = 0, Y_{i2} = 0 | X_i) = S_{i1} S_{i2} + \rho \sqrt{F_{i1} S_{i1} F_{i2} S_{i2}}.$$

We maximise the following (unpenalised) log-likelihood to estimate the parameters $\boldsymbol{\beta}_1, \boldsymbol{\beta}_2$ and $\rho$:

$$l(\boldsymbol{\beta}_1, \boldsymbol{\beta}_2, \rho;) = \sum_{i=1}^{N} y_{i1} y_{i2} \log(p_{11i}) + y_{i1}(1 - y_{i2}) \log(p_{10i}) + (1 - y_{i1}) y_{i2} \log(p_{01i}) + (1 - y_{i1})(1 - y_{i2}) \log(p_{00i}).$$

where $p_{abi} = P(Y_{i1} = a, Y_{i2} = b | X_i)$. There are no closed-form solutions to maximise the derivatives of this log-likelihood and so numerical optimisation is required. At the time of prediction, the joint and marginal probabilities of each outcome can be obtained directly for each new individual.

Of note is that the residual correlation parameter, $\rho$, is constrained by the marginal probabilities, and cannot therefore take values in the full $[-1,1]$ range [30]; previous studies have shown this to impact the degrees of dependency between the outcomes that the approach can handle [32,33].

## 3.4 Multivariate Bayesian Probit CPM

Our final approach follows naturally from the case of modelling multiple continuous outcomes through multivariate linear regression [5]. Limiting again to the case for $K = 2$ for ease of exposition, let $Z_{i1}$ and $Z_{i2}$ denote two latent variables for each outcome, such that

$$Y_{ij} = \begin{cases} 1 & \text{if } Z_{ij} > 0 \\ 0 & \text{if } Z_{ij} \leq 0 \end{cases}$$

where $Z_{ij} = \beta_{0,j} + X_i \beta_j + \epsilon_{ij}$. Dependency between the outcomes is induced by assuming a joint distribution on $\epsilon_{ij}$ for $j = 1,2$. Given the benefits of specifying dependency through a multivariate normal distribution, we propose to fit a multivariate probit model on $Y$, by assuming

$$\begin{pmatrix} \epsilon_1 \\ \epsilon_2 \end{pmatrix} \sim N\left(\begin{pmatrix} 0 \\ 0 \end{pmatrix}, \begin{pmatrix} 1 & \rho \\ \rho & 1 \end{pmatrix}\right).$$

We propose to fit this model using Bayesian inference, since parameter estimation can be obtained naturally using MCMC methods (which is necessary for $K > 2$). In this study we set the prior distributions to $\rho \sim \text{Unif}(-1,1)$ and

$$\begin{pmatrix} \boldsymbol{\beta_1} \\ \boldsymbol{\beta_2} \end{pmatrix} \sim N\left(\begin{pmatrix} \mathbf{0} \\ \mathbf{0} \end{pmatrix}, \begin{pmatrix} \boldsymbol{\Sigma_1} & \boldsymbol{\Sigma_1 \Sigma_2} \\ \boldsymbol{\Sigma_1 \Sigma_2} & \boldsymbol{\Sigma_2} \end{pmatrix}\right).$$

where $\boldsymbol{\Sigma_1 \Sigma_2} = \mathbf{0}$ and $\boldsymbol{\Sigma_j} = \text{diag}(10)$, for $j = 1,2$. In most medical applications, it might be more appropriate to set $\rho \sim \text{Unif}(0,1)$, since the correlation would usually be positive (or could be constructed to be such through transformation of the outcomes prior to modelling) [34].

At time of prediction for a new individual with covariates $X_i^*$, estimates of $P(Y_{i1} = 1, Y_{i2} = 1 | X_i^*)$ can be obtained through the cumulative distribution function of the bivariate standard normal distribution, $\Phi$, as $\Phi(X_i^* \boldsymbol{\beta_1}, X_i^* \boldsymbol{\beta_2}, \rho)$. Similarly, estimates of $P(Y_{i1} = 1, Y_{i2} = 0 | X_i^*)$ and $P(Y_{i1} = 0, Y_{i2} = 1 | X_i^*)$ can be obtained through $\Phi(X_i^* \boldsymbol{\beta_1}, -X_i^* \boldsymbol{\beta_2}, -\rho)$ and $\Phi(-X_i^* \boldsymbol{\beta_1}, X_i^* \boldsymbol{\beta_2}, -\rho)$, respectively.

In this study, we implemented this model using JAGS (Just Another Gibbs Sampler), using the R package rjags [35]. We took 10000 posterior samples of each parameter and summarised them over the final 5000 samples (i.e. 5000 burn-in).

## 4 Simulation Study

### 4.1 Aim

We designed a simulation study to investigate the effects on predictive performance of developing CPMs that model multiple outcomes using the aforementioned approaches, compared with modelling each outcome separately through univariate analyses. We designed and report the simulation in line with best practice [36].

### 4.2 Data-generating mechanisms

Throughout all simulations we assume that we have an IPD that includes 5000 individuals, on which one is interested in developing CPMs for two binary outcomes of interest. The IPD includes two continuous covariates that we generate as $X_1 \sim N(0,1)$ and $X_2 \sim N(0,1)$. Additionally, each observation within the IPD has two (potentially dependent) binary outcomes, $Y_1$ and $Y_2$, which we simulate according to established methods [37,38]. Specifically, we generated the binary outcomes such that the marginal probabilities satisfied

$$P(Y_{i1} = 1|X_i) = \left[1 + \exp\left(-(\beta_{01} + \beta_{11}X_{i1} + \beta_{21}X_{i2})\right)\right]^{-1},$$

$$P(Y_{i2} = 1|X_i) = \left[1 + \exp\left(-(\beta_{02} + \beta_{12}X_{i1} + \beta_{22}X_{i2})\right)\right]^{-1},$$

where we fixed $\boldsymbol{\beta}_1 = (\beta_{01}, \beta_{11}, \beta_{21}) = (-1, \log(2), 0)$ and $\boldsymbol{\beta}_2 = (\beta_{02}, \beta_{12}, \beta_{22}) = (-1.5, 0, \log(3))$, meaning that $X_1$ and $X_2$ only predicted $Y_1$ and $Y_2$, respectively. These coefficient values give (marginal) outcome proportions of 29% and 23% for $Y_1$ and $Y_2$, respectively. We also considered a sensitivity analysis where $\beta_{01}$ was set to -3 and $\beta_{02}$ was set to -3.5, which results in lower (marginal) outcome proportions of 6% and 5% for $Y_1$ and $Y_2$, respectively. Dependency between the outcomes was induced (while satisfying the above marginal probabilities) by generating two latent variables $Z_{i1}$ and $Z_{i2}$, from a multivariate standard normal distribution, with correlation $\rho$. We then applied a probability transform such that $\epsilon_{i1} = \text{logit}(\Phi(Z_{i1}))$ and $\epsilon_{i2} = \text{logit}(\Phi(Z_{i2}))$, where $\text{logit}(.)$ is the inverse logistic function, and $\Phi(.)$ is the cumulative distribution function of the standard normal distribution. The two binary outcomes were then generated such that

$$Y_{ij} = I\left(\epsilon_{ij} \leq (\beta_{0j} + \beta_{1j}X_{i1} + \beta_{2j}X_{i2})\right); j = 1,2$$

with $I(.)$ being the indicator function. Within this data generating process, $\rho$ controls the level of dependence (after conditioning on the covariates) between the two outcomes; the outcomes are conditionally independent when $\rho = 0$. Hence, $\rho$ was varied across simulation scenarios through values of $\{0, 0.25, 0.50, 0.75, 0.95\}$. Note that for $\rho > 0$ the correlation between $Y_1$ and $Y_2$ is less than $\rho$ due to the non-linear probability transform applied to $Z_1$ and $Z_2$ (see, e.g., [39]).

Finally, to test the predictive performance of all the analysis methods (outlined below), we generated an independent set of 10000 observations, to serve as a validation set. This was generated using the same data-generating mechanisms as the IPD, thereby representing internal validation.

### 4.3 Methods considered
Within each generated IPD set, we fit the following analysis models using maximum likelihood estimation or MCMC, as appropriate: (i) two independent CPMs, one for each outcome (section 2.2), (ii) stacked regression (section 2.3), (iii) probabilistic classification chains (section 3.1), (iv) multinomial logistic regression (section 3.2), (v) multivariate logistic regression (section 3.3), and (vi) multivariate Bayesian probit (section 3.4). All analysis models included both $X_1$ and $X_2$ (i.e. no variable selection).

### 4.4 Target Predictions
The main target outputs/predictions of interest were the predicted marginal probabilities $P(Y_{1i} = 1)$ and $P(Y_{2i} = 1)$, along with the following predicted joint probabilities: $P(Y_{1i} =$

$1, Y_{2i} = 1$), $P(Y_{1i} = 1, Y_{2i} = 0)$, and $P(Y_{1i} = 0, Y_{2i} = 1)$, where the conditionals on $\boldsymbol{X}_1$ and $\boldsymbol{X}_2$ have been omitted for brevity. Details of how each method estimates these joint and marginal probabilities were described in sections **Error! Reference source not found.** and 3.

### 4.5 Performance measures

The data-generating mechanisms were repeated across 100 iterations for all simulation scenarios (i.e. for all values of $\rho$). We used the validation sets, generated separately in each iteration, to assess the CPMs in terms of calibration (agreement between the expected event rate and the observed event rate, across the full risk range), discrimination (ability of the model to separate cases from controls) and mean squared error of the predicted risks compared with the 'true' (data-generating) risks. The mean squared error (MSE) was defined as $n^{-1} \sum_{i=1}^{n}(\pi_i - \hat{\pi}_i)^2$, where $n = 10000$ in the validation set, $\pi_i$ is the data-generating risk for observation $i$ (either joint or marginal depending on the MSE being calculated) and $\hat{\pi}_i$ is the corresponding estimated risk from each model. Calibration was quantified with the calibration-in-the-large (ideal value 0) and slope (ideal value 1). For the marginal outcomes, this was estimated by fitting a logistic regression model in the validation set for each observed marginal outcome and with the logit of the estimated marginal risk (from each model) as the only covariate; this covariate was used as an offset for estimation of calibration-in-the-large [40]. The calibration of the joint outcomes was estimated using multinomial methods, as previously described in detail [41,42] (see supplementary methods for details). Discrimination of all models was estimated with the area under the receiver operating characteristic curve (AUC). All performance measures were summarised across the 100 iterations.

### 4.6 Software

The simulation was implemented in R version 3.6.1 [43], along with the following packages: tidyverse [44], pROC [45], rjags [35], coda [46], pbivnorm [47], glmnet [48], VGAM [49–51] and nnet [29]. The code was written by the lead author and is available on GitHub at https://github.com/GlenMartin31/Multivariate-Binary-CPMs.

### 4.7 Simulation Results

We here present the simulation results for the case where the marginal outcome proportions were 29% and 23% for $\boldsymbol{Y}_1$ and $\boldsymbol{Y}_2$, respectively. Quantitatively similar results were found for the sensitivity analysis where the marginal outcome proportions were lowered to 6% and 5% for $\boldsymbol{Y}_1$ and $\boldsymbol{Y}_2$, respectively (simulation results available through the GitHub page).

**Table 1** shows the empirical relationships between $\rho$ (a key simulation parameter) and the correlation between the binary outcomes (as defined by the Phi coefficient). The observed joint probability of both outcomes (averaged across all iterations) ranged from 6.5% for $\rho = 0$ to 16.1% for $\rho = 0.95$.

When $\rho = 0$, all models returned estimates of the overall joint probabilities that were calibrated well with observed probabilities in the validation data (**Figure 1**). However, as $\rho$ increased, the calibration-in-the-large for $P(Y_{1i} = 1, Y_{2i} = 1)$ increased above 0 for univariate CPMs and stacked regression, indicating that these models (which ignore conditional dependency of the outcomes) underestimate the joint risk. In contrast, for all methods that account for dependence in the outcomes (i.e. probabilistic classification chains, multinomial logistic regression, multivariate logistic regression and multivariate Bayesian probit regression), the calibration-in-the-large were consistently close to 0. However, for higher values of $\rho$, the calibration-in-the-large deviated from 0 for multivariate logistic regression, especially when estimating $P(Y_{1i} = 0, Y_{2i} = 1)$; this is as expected as discussed in section 3.3. In terms of marginal outcome risk, the calibration-in-the-large was sufficiently close to 0 for all methods across all values of $\rho$, indicating the overall expected marginal event rates matched the observed marginal event rates for all methods, as expected (**Supplementary Figure 1**).

Similar findings were observed for the calibration slope for both the joint outcome risk (**Figure 2**) and marginal outcome risk (**Supplementary Figure 2**). Specifically, predicted risks for each outcome were well calibrated for the two observed marginal probabilities for all models (**Supplementary Figure 2**). For the joint outcome probabilities, only probabilistic classification chains, multinomial logistic regression and multivariate Bayesian probit regression (i.e. the methods that account for dependence in the outcomes across the full range of $\rho$) had a calibration slope close to 1 for increasing $\rho$ (**Figure 2**). Multivariate logistic regression was miscalibrated for predicting $P(Y_{1i} = 0, Y_{2i} = 1)$ for $\rho > 0.5$.

Discrimination was consistent across all methods and all simulation scenarios for both joint outcomes (**Figure 3**) and marginal outcomes (**Supplementary Figure 3**). Upon validation, the AUC was slightly higher for probabilistic classification chains, multinomial logistic regression, and multivariate Bayesian probit regression CPMs in terms of predicting the joint probabilities for the highest values of $\rho$, but differences were modest.

Finally, the MSE showed that all the methods consistently estimated the risks of the marginal outcomes (**Supplementary Figure 4**), but the CPMs developed using probabilistic classification chains, multinomial logistic regression and multivariate Bayesian probit regression had lower (i.e. better) MSE for the joint outcome risks as $\rho$ increased compared with the methods that ignored the conditional dependency of the outcomes or multivariate logistic regression (**Figure 4**).

# 5 Empirical Study

## 5.1 Data Source, Study Population and Outcomes

Data were obtained from the Medical Information Mart for Intensive Care III (MIMIC-III), which contains freely-available and de-identified critical care data from the Beth Israel Deaconess Medical Center in Boston, Massachusetts, between 2001 and 2012 [52].

For the purposes of this empirical study of the methods, we considered the prediction of a binary indication of acute kidney injury (AKI) occurring within 48 hours after an ICU admission, and a binary indication of a total length of stay (LOS) on ICU of over 5 days. In this study, we defined an ICU admission to be any that lasted at least 24 hours, and we took the end of day 1 on ICU as our prediction time (the time point at which a prediction is made); hence, the AKI outcome was determined at the end of day 3. The LOS outcome aims to capture overall ICU severity, and matches previous studies in this area [53]. Ideally, LOS should be analysed as a continuous or count outcome, but we considered LOS as binary here for illustrative purposes.

We extracted a cohort of patients over 18 years of age from the MIMIC-III database who were admitted to ICU for any cause for at least 24 hours. We extracted information on patients' age, gender, ethnicity, type of admission, and vital signs and lab results over the first 24 hours of each ICU admission (summarised as minimum, mean and maximum values).

To define AKI, we extracted the maximum creatinine value for each patient collected between 24 and 72 hours after initial ICU admission. AKI was defined as present (coded as 1) if the maximum creatinine within 48 hours after the prediction time was either: (i) more than 1.5 times the minimum day 1 creatinine value, or (ii) over 0.3mg/dL greater than the minimum day 1 creatinine value; AKI was coded 0 (absent) otherwise. This definition follows published clinical guidance [54].

We excluded any ICU admission with indication of reduced kidney function within the first 24 hours, by excluding those with an estimated glomerular filtration rate GFR (eGFR) less than 60 mL/min/1.73 $m^2$ at baseline. The eGFR was calculated by the MDRD study equation [55], using each patient's minimum creatinine value within the first 24 hours. Patients with missing outcomes or who died within the hospitalisation of their ICU stay were also excluded; this was for simplicity since this application was only for illustration. We only included a patient's first ICU admission for a given hospitalisation.

We developed CPMs on the extracted cohort using the methods outlined in section 2 and section 3 with the aim of estimating the marginal and joint probabilities of AKI and total ICU LOS ≥ 5 days. All the models included identical predictors (**Supplementary Table 1**), and therefore differed only in their estimation/ inference processes. To emphasise, the purpose was not to derive a new CPM for use in clinical practice, but rather to illustrate and apply the proposed analytical methods using real-world clinical data. We evaluated each of the models in terms of their respective calibration and discrimination metrics, using the same techniques as described in section 4.5. The large sample means that overfitting was not a concern [19]; nonetheless, we report predictive performance in a random hold-out sample (30%), simply to have some unused data to check the models in, rather than relying on development data alone (noting that split-sample method is not preferred in practice [1–3]).

Missing data in any predictor variables was imputed using multiple imputation, where we generated 20 imputed datasets [56]. The imputation models included all of the covariates, plus the two outcomes [57]. Within each imputed dataset, the CPMs were developed using each analytical method, which were then applied to the hold-out test samples to estimate the calibration and discrimination of each model. Performance metrics upon validation were then pooled across the imputations using Rubin's rules [56].

All data extraction was performed using an SQL script written by the lead author (which is available on Github: https://github.com/GlenMartin31/Multivariate-Binary-CPMs). Analysis was performed using R version 3.6.1 [43], along with the packages stated in section 4.6.

## 5.2 Empirical Study Results

A total of 24459 ICU admissions were included in the analysis; **Table 2** presents an overview of a baseline summary of the whole cohort extracted from the MIMIC-III database. The correlation between the outcomes was 0.08 (Phi coefficient), with an observed joint probability of both outcomes being 4.46%. The marginal probability of long LOS was 20.3% and for AKI was 16.1%.

**Figure 5** shows the calibration and discrimination (in the hold-out sample) for each method in terms of estimating the joint outcome risks. As with the simulation study, methods that account for dependence in the outcomes (i.e. probabilistic classification chains, multinomial logistic regression, multivariate logistic regression and multivariate Bayesian probit regression) were well calibrated for all outcomes, with calibration-in-the-large and calibration slope close to 0 and 1, respectively. The models that do not account for outcome dependency significantly under-predicted the joint outcome risk, with a calibration-in-the-large over 0. All models had similar discrimination. These results align with those from the simulation study; here, we have an observed correlation in the outcomes of 0.08, which approximately corresponds to the simulation scenario where $\rho = 0.25$ (see **Table 1**).

## 6 Discussion

This study presents four methods for developing CPMs that respect the dependence between multiple clinical outcomes. As expected, only the methods that condition on each outcome (probabilistic classification chains and multinomial logistic regression) or model the correlation explicitly (multivariate logistic regression and multivariate Bayesian probit regression) provide reliable estimates of joint risks. All methods had similar predictive performance in terms of predicting the marginal risks of each outcome.

There has been little previous research published on developing CPMs that aim to predict multiple outcomes simultaneously [18]. Most CPMs have been developed for individual or composite outcomes. However, as multi-morbidity becomes a priority for health services around the world, there is a pressing need for developing methods to more accurately predict the risk of developing multiple conditions – in both clinical decision-making and service planning. Traditional approaches (e.g. Charlson Comorbidity Index [58]) involve rudimentary metrics, assigning crude weights to different conditions that cannot predict the co-occurrence of outcomes. This study shows that accurate prediction of joint outcome risks, which are required to fully support multi-morbidity care planning, is achieved by developing CPMs that account for dependence in the outcomes. Models that do not account for outcome dependence underestimate the joint outcome risks, thereby underestimating multi-morbidity risk if used in this context. While this is intuitive from a statistical perspective, CPMs are usually not developed accordingly [4,9,10,18]. As such, the findings from this study have implications for multi-outcome risk prediction, which is becoming an increasing priority. Advantageously, all the methods proposed can be implemented using standard statistical software (although multivariate logistic regression and multivariate Bayesian probit regression do require user coding), meaning they could be readily applied to develop real-world CPMs for multi-outcome prediction.

Nonetheless, we note that all CPMs are developed with a particular prediction task in mind, and as such not all CPMs will aim – or indeed need – to accurately estimate joint risks of multiple outcomes. This study shows that all approaches accurately predicted the risks of each outcome individually, meaning the models in section 3 are still useful at predicting marginal risk. It is important to emphasise that the methods that utilise data from multiple outcomes can also leverage the information contained across outcomes, with the associated advantages that this brings. For example, such advantages have been widely shown in the joint modelling literature where one can improve prediction of time-to-event outcomes through the information from repeated biomarkers [59,60], and also in cross-sectional data of correlated binary outcomes [4,9,10].

Here we focus on predicting binary outcomes, but continuous, ordinal and time-to-event outcomes are also commonly required from CPMs [3]. While most of the methods described in this paper generalise naturally to predicting continuous outcomes, further consideration will be required for ordinal and time-to-event data. Explicitly, for time-to-event, one should account for competing risks (e.g. death), especially in the case of predicting multi-morbidity where outcomes occur over several years. Multi-state survival models present a way of doing this [11], and have been used to develop CPMs to predict risk of moving between disease/condition/pathway states (e.g. [12]). For example, previous work has investigated the risk of developing cardiovascular disease across strata of different predictors, while acknowledging competing risks due to death [61]. However, the use of multi-state models to develop CPMs for multi-morbidity prediction is not commonplace and would require methodological developments due to combinatorial complexities arising by the number of states (i.e. outcome combinations). As such, further research is required to extend the approaches in this paper to consider time-to-event CPMs through a multi-state and competing risk framework; some of the methods outlined here might provide a foundation to doing this.

Similarly, a patient's care pathway comprises a mixture of outcome 'types'; thus, it would be advantageous if the methods to develop CPMs for multiple outcomes could handle this heterogeneity. For example, one might need to predict continuous (e.g. blood test), binary (e.g. procedural complication) and time-to-event (e.g. time-to-readmission) outcomes simultaneously. While stacked regression provides a natural way of doing this by operating on the linear predictor scale (as discussed in section 2.3), this method relies on conditional independence so cannot estimate joint outcome risks. In contrast, the multivariate probit model has been used to model continuous and binary outcomes simultaneously by correlating the error terms of a linear model with the latent error term for the binary outcome [10,62]. The use of copula methods within the multivariate probit model might be one way to generalise this approach to work for any pair of outcome types [63]. Further research is warranted to explore this, and to consider the extension of the other methods to handle different outcomes.

### 6.1   Limitations

Several limitations should be considered for this study. First, we have only considered binary outcomes where the occurrence of one outcome does not prevent the occurrence of the others. In practice, competing risks will need to be accounted for and the biases incurred by failing to consider this were not explored here. Second, we only generated two predictors in the simulation study; we acknowledge that most CPMs include more than two covariates, but one could regard the two simulated predictors as summaries of several variables. Similarly, neither the simulation nor the critical care example considered variable selection. Variable selection might be especially important where different predictors are associated with different outcomes, or where the direction of the association of a given predictor differs across outcomes. Considering variable selection and heterogeneity in associations should be considered for future work. Third, we only considered big data examples and simulations with large numbers of observations; therefore, further research is needed to explore the concepts of this paper in settings where overfitting might be a concern (e.g. penalisation) [19]. Fourth, the simulations and real-world example only considered two outcomes, and computational cost might rise if aiming to predict more outcomes simultaneously. Finally, we only considered methods embedded within a regression framework, and we acknowledge that we have not considered machine learning, classification-based approaches such as multi-label neural networks [17]. Nonetheless, interpretable, machine learning multivariate CPMs are a grand challenge.

## 6.2 Conclusion

This paper reports four approaches that can advance CPMs beyond the current disconnected prediction of single conditions to combinatorial approaches that reflect the real-world challenge of multi-morbidity. Any CPM that aims to predict joint risk of multiple outcomes should only be based on methods that explicitly model the correlation structure. In such a situation, our results suggest that probabilistic classification chains, multinomial logistic regression or the multivariate probit model might be the most appropriate choice. Approaches that model outcome dependency more accurately reflect real-world healthcare and benefit from the well-known advantages to inference that analysing multiple outcomes simultaneously offers.

# 8 Tables

**Table 1**: Empirical results of the correlation between $Y_1$ and $Y_2$ for each value of $\rho$ in the simulation, along with the observed joint outcome event rates

| $\rho$ | $\text{Corr}(Y_1, Y_2)$ | $P(Y_1 = 1, Y_2 = 1)$ | $P(Y_1 = 1, Y_2 = 0)$ | $P(Y_1 = 0, Y_2 = 1)$ |
|---|---|---|---|---|
| 0 | 0.000 | 0.065 | 0.222 | 0.162 |
| 0.25 | 0.110 | 0.087 | 0.201 | 0.142 |
| 0.50 | 0.233 | 0.110 | 0.178 | 0.118 |
| 0.75 | 0.371 | 0.136 | 0.152 | 0.092 |
| 0.95 | 0.503 | 0.161 | 0.126 | 0.067 |

**Table 2**: Baseline summary of the patient demographics, characteristics and lab/vital results over the first 24 hours of an intensive care unit (ICU) admission, for the whole cohort.

| Characteristic | Summary | Missing Data, n (% of cohort) |
|---|---|---|
| N | 24459 | |
| **Demographics** | | |
| Age, mean (min, max) | 60.92 (18.02, 99.28) | 0 (0%) |
| Age group, n (%) | | 0 (0%) |
|   <30 | 1406 (5.75%) | |
|   30-40 | 1534 (6.27%) | |
|   40-50 | 3091 (12.6%) | |
|   50-60 | 4958 (20.3%) | |
|   60-70 | 5522 (22.6%) | |
|   70-80 | 4616 (18.9%) | |
|   >80 | 3332 (13.6%) | |
| Male, n (%) | 14438 (59.0%) | 0 (0%) |
| Admission Type, n (%) | | 0 (0%) |
|   Elective | 4627 (18.9%) | |
|   Urgent | 609 (2.49%) | |
|   Emergency | 19223 (78.6%) | |
| Ethnicity, n (%) | | 2665 (10.9%) |
|   White | 17637 (72.1%) | |
|   Asian | 593 (2.42%) | |
|   Black | 1998 (8.17%) | |
|   Hispanic | 867 (3.54%) | |
|   Other | 709 (2.90%) | |
| **Lab Tests – summary over first 24 hours on ICU** | | |
|   Mean bicarbonate, mean (min, max) | 24.4 (8.00, 51.5) | 123 (0.50%) |
|   Mean creatinine, mean (min, max) | 0.82 (0.10, 4.73) | 0 (0%) |
|   Mean chloride, mean (min, max) | 105.2 (64.5, 142.0) | 85 (0.35%) |
|   Mean haemoglobin, mean (min, max) | 11.1 (3.33, 19.9) | 59 (0.24%) |
|   Mean platelet count, mean (min, max) | 224.6 (7.50, 1646.2) | 84 (0.34%) |
|   Mean potassium, mean (min, max) | 4.09 (2.30, 8.70) | 2 (0.01%) |
|   Mean partial thromboplastin time, mean (min, max) | 35.3 (14.4, 150.0) | 2596 (10.6%) |
|   Mean international normalised ratio, mean (min, max) | 1.37 (0.50, 18.2) | 2537 (10.4%) |
|   Mean prothrombin time, mean (min, max) | 14.9 (8.00, 131.1) | 2542 (10.4%) |
|   Mean white blood cell count, mean (min, max) | 11.8 (0.10, 247.9) | 137 (0.56%) |
| **Vital signs – summary over first 24 hours on ICU** | | |
| Mean heart rate, mean (min, max) | 86.2 (31.2, 155.0) | 203 (0.83%) |
| Mean systolic blood pressure, mean (min, max) | 119.0 (74.1, 203.0) | 220 (0.90%) |
| Mean diastolic blood pressure, mean (min, max) | 61.9 (27.4, 127.0) | 220 (0.90%) |
| Mean respiration rate, mean (min, max) | 18.5 (8.00, 41.8) | 225 (0.92%) |
| Mean temperature (Celsius), mean (min, max) | 36.9 (32.6, 39.8) | 703 (2.87%) |
| Mean oxygen saturation, mean (min, max) | 97.5 (73.5, 100.0) | 211 (0.86%) |
| Mean glucose, mean (min, max) | 135.8 (52.0, 661.8) | 299 (1.22%) |
| **Outcomes** | | |
| Total ICU Length of Stay ≥ 5 days, n (%) | 4957 (20.3%) | 0 (0%) |
| Acute Kidney Injury by day 3 on ICU, n (%) | 3930 (16.1%) | 0 (0%) |

# 9 Figures

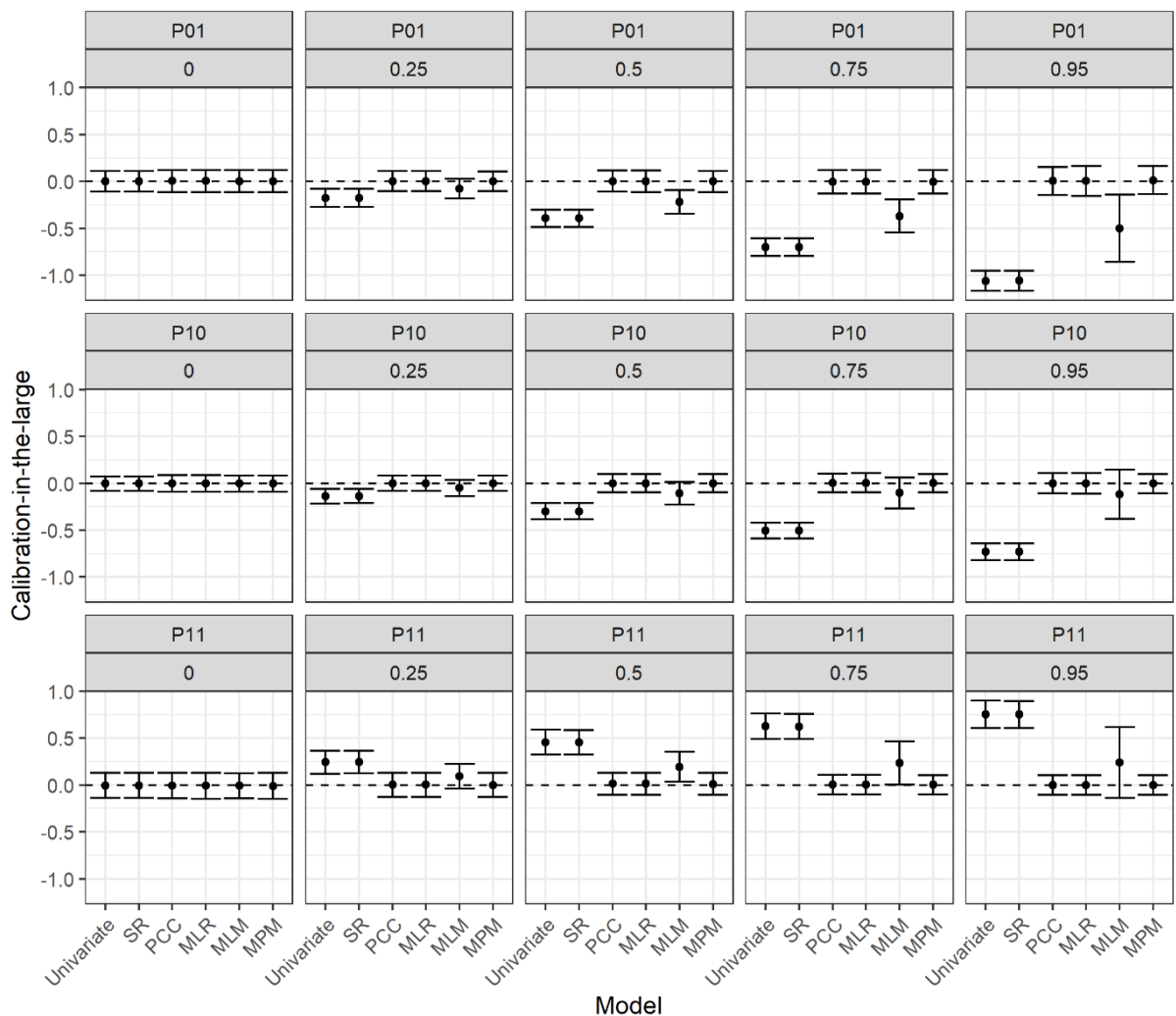

**Figure 1**: Calibration-in-the-large for each model across all simulation scenarios, upon validation. Each column of plots corresponds to a value of $\rho$, while each row of plots is a joint outcome as follows: $P11$ denotes $P(Y_{1i} = 1, Y_{2i} = 1)$, $P10$ denotes $P(Y_{1i} = 1, Y_{2i} = 0)$, and $P01$ denotes $P(Y_{1i} = 0, Y_{2i} = 1)$. The dashed horizontal lines show the reference value for the calibration-in-the-large of 0. The models are as follows and as described in the methods section: Univariate = two independent CPMs, SR = stacked regression, PCC = probabilistic classification chains, MLR = multinomial logistic regression, MLM = multivariate logistic model, and MPM = multivariate Bayesian probit model.

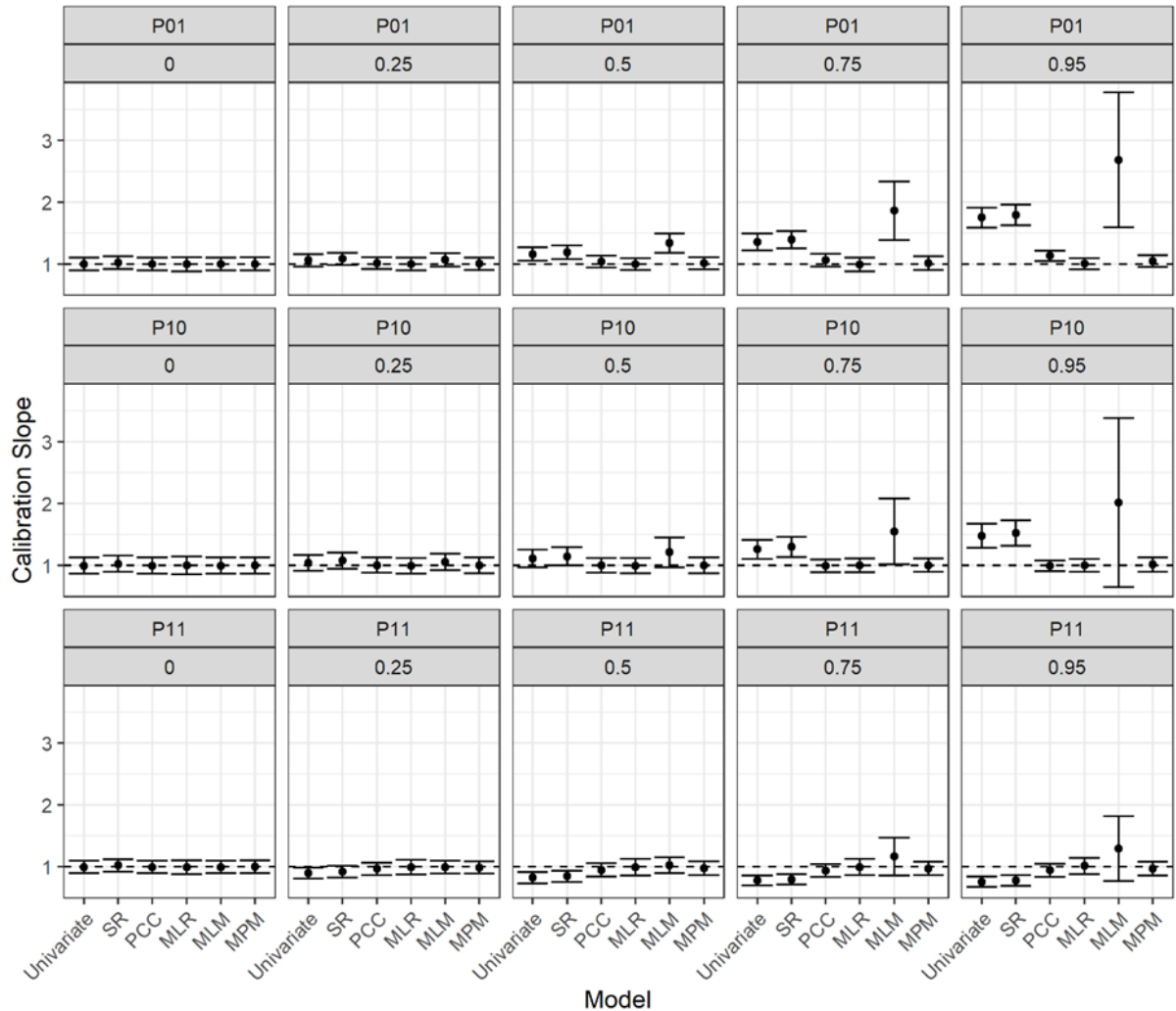

**Figure 2:** Calibration slope for each model across all simulation scenarios, upon validation. Each column of plots corresponds to a value of $\rho$, while each row of plots is a joint outcome as follows: $P11$ denotes $P(Y_{1i} = 1, Y_{2i} = 1)$, $P10$ denotes $P(Y_{1i} = 1, Y_{2i} = 0)$, and $P01$ denotes $P(Y_{1i} = 0, Y_{2i} = 1)$. The dashed horizontal lines show the reference value for the calibration slope of 1. The models are as follows and as described in the methods section: Univariate = two independent CPMs, SR = stacked regression, PCC = probabilistic classification chains, MLR = multinomial logistic regression, MLM = multivariate logistic model, and MPM = multivariate Bayesian probit model.

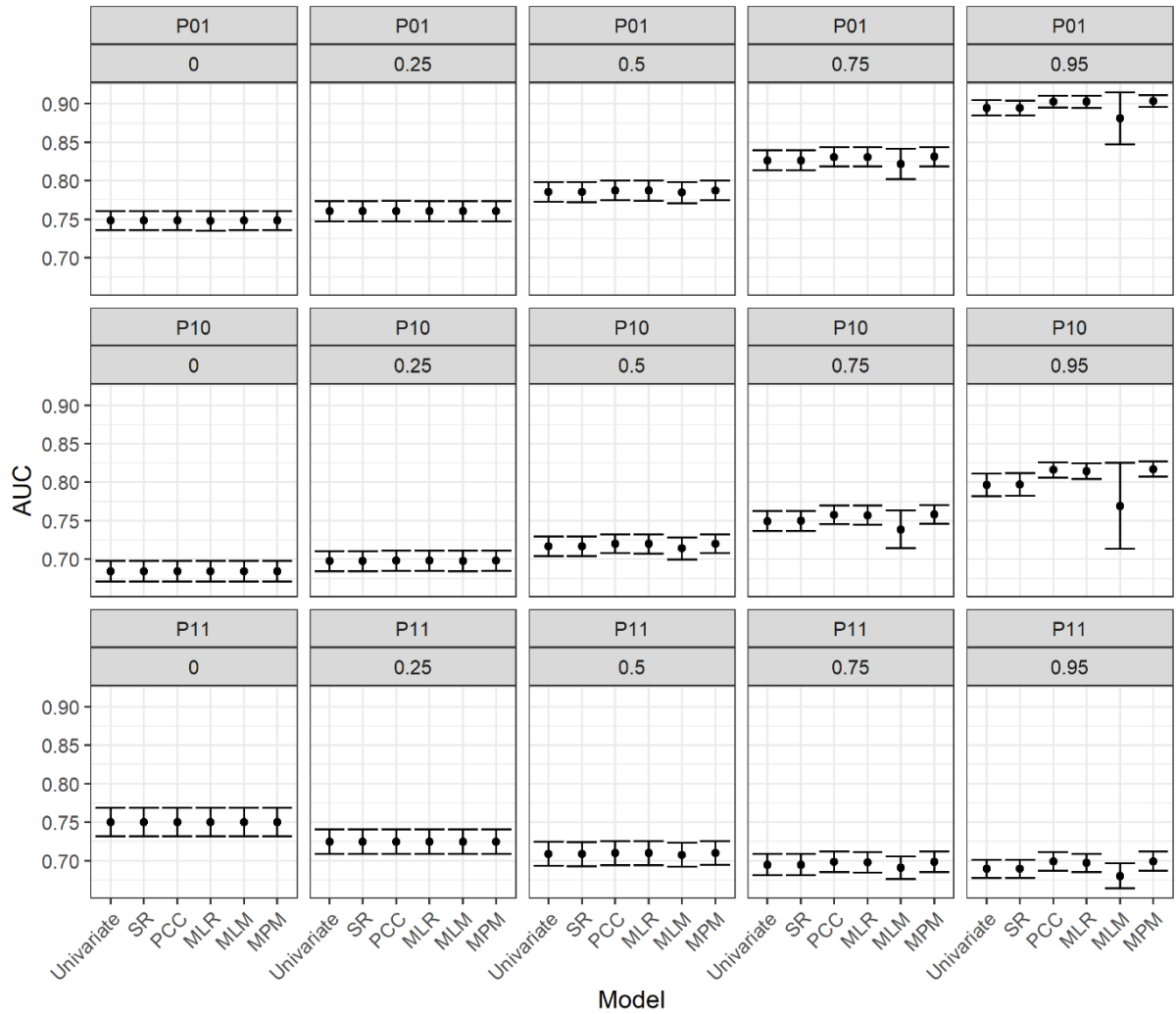

**Figure 3:** Area under receiver operating characteristic curve (AUC) for each model across all simulation scenarios, upon validation. Each column of plots corresponds to a value of $\rho$, while each row of plots is a joint outcome as follows: $P11$ denotes $P(Y_{1i} = 1, Y_{2i} = 1)$, $P10$ denotes $P(Y_{1i} = 1, Y_{2i} = 0)$, and $P01$ denotes $P(Y_{1i} = 0, Y_{2i} = 1)$. The models are as follows and as described in the methods section: Univariate = two independent CPMs, SR = stacked regression, PCC = probabilistic classification chains, MLR = multinomial logistic regression, MLM = multivariate logistic model, and MPM = multivariate Bayesian probit model.

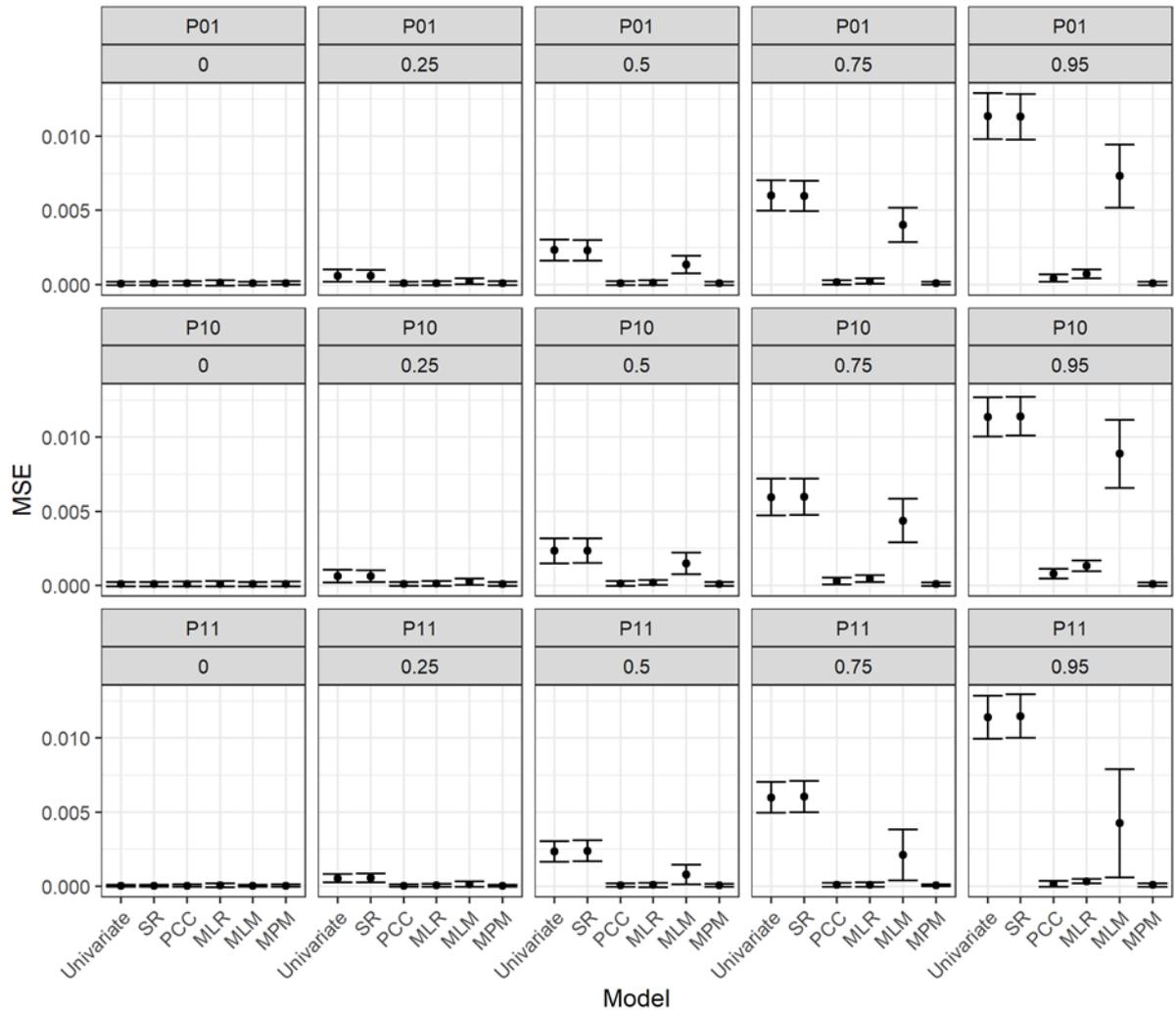

**Figure 4**: Mean squared error (MSE) for each model across all simulation scenarios (lower is better), upon validation. Each column of plots corresponds to a value of $\rho$, while each row of plots is a joint outcome as follows: $P11$ denotes $P(Y_{1i} = 1, Y_{2i} = 1)$, $P10$ denotes $P(Y_{1i} = 1, Y_{2i} = 0)$, and $P01$ denotes $P(Y_{1i} = 0, Y_{2i} = 1)$. The models are as follows and as described in the methods section: Univariate = two independent CPMs, SR = stacked regression, PCC = probabilistic classification chains, MLR = multinomial logistic regression, MLM = multivariate logistic model, and MPM = multivariate Bayesian probit model.

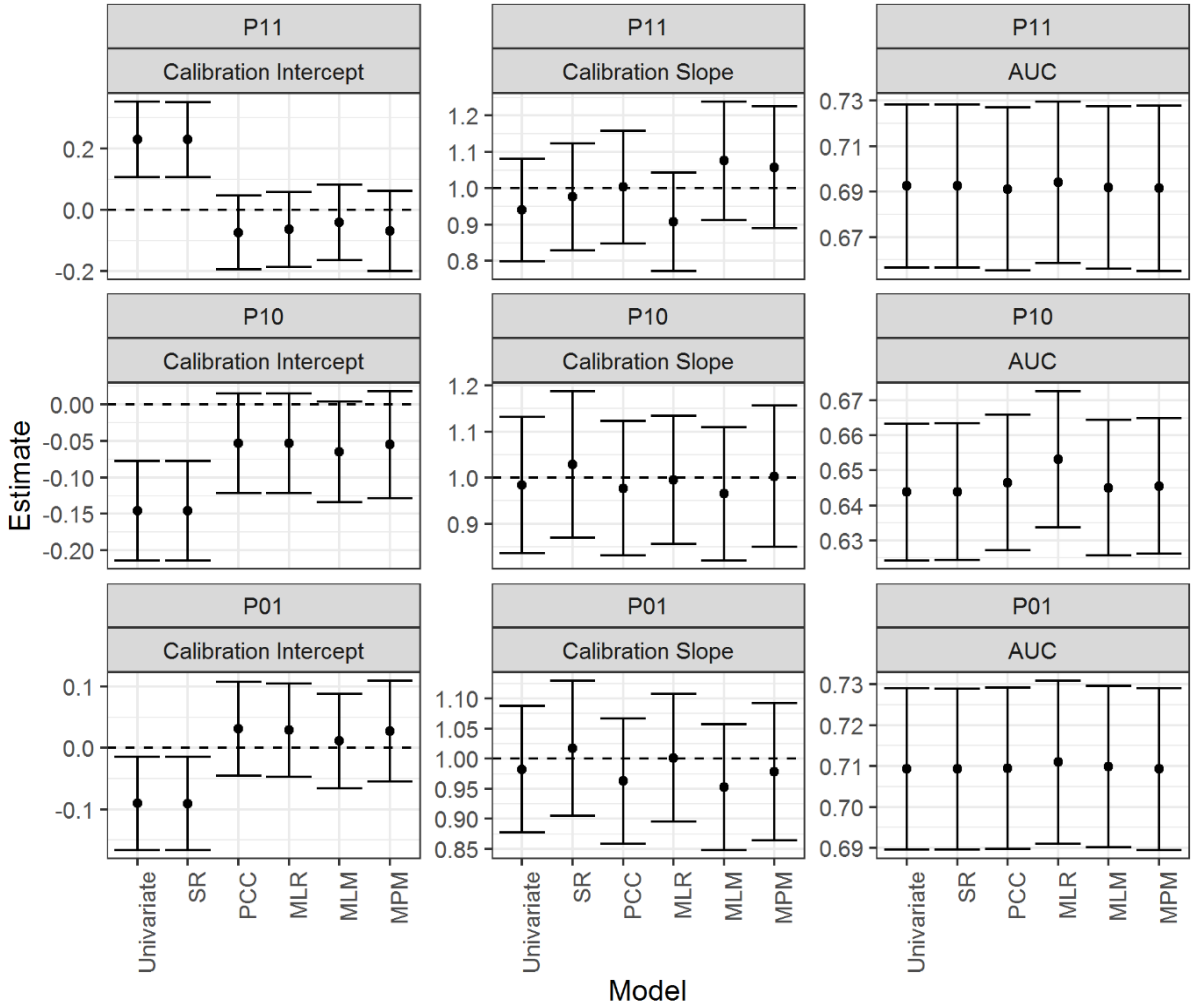

**Figure 5**: Internal validation (hold-out sample) performance results for each model in the MIMIC-III dataset. Each column of plots corresponds to calibration-in-the-large, calibration slope and AUC, while each row of plots is a joint outcome as follows: $P11$ denotes $P(ICU > 5_i = 1, AKI_i = 1)$, $P10$ denotes $P(ICU > 5_i = 1, AKI_i = 0)$, and $P01$ denotes $P(ICU > 5_i = 0, AKI_i = 1)$. The models are as follows and as described in the methods section: Univariate = two independent CPMs, SR = stacked regression, PCC = probabilistic classification chains, MLR = multinomial logistic regression, MLM = multivariate logistic model, and MPM = multivariate Bayesian probit model.

# 10 Supplementary Methods

We here describe our approach to calibration of the joint outcomes. Each of the joint outcome combinations form a multinomial outcome; specifically, for $K = 2$ we have the following joint outcome combinations: $\{Y_1 = 1, Y_2 = 1\}$, $\{Y_1 = 1, Y_{2i} = 0\}$, $\{Y_1 = 0, Y_2 = 1\}$ and $\{Y_1 = 0, Y_2 = 0\}$. The calibration of the prediction models at estimating each joint outcome risk is based on fitting multinomial logistic regression models of the form

$$\begin{cases} \log\left(\frac{P(Y_{i1} = 1, Y_{i2} = 1)}{P(Y_{i1} = 0, Y_{i2} = 0)}\right) = \alpha_{11} + \alpha_{12}\log\left(\frac{\widehat{P_{11}}}{\widehat{P_{00}}}\right) + \alpha_{13}\log\left(\frac{\widehat{P_{10}}}{\widehat{P_{00}}}\right) + \alpha_{14}\log\left(\frac{\widehat{P_{01}}}{\widehat{P_{00}}}\right) \\ \log\left(\frac{P(Y_{i1} = 1, Y_{i2} = 0)}{P(Y_{i1} = 0, Y_{i2} = 0)}\right) = \alpha_{21} + \alpha_{22}\log\left(\frac{\widehat{P_{11}}}{\widehat{P_{00}}}\right) + \alpha_{23}\log\left(\frac{\widehat{P_{10}}}{\widehat{P_{00}}}\right) + \alpha_{24}\log\left(\frac{\widehat{P_{01}}}{\widehat{P_{00}}}\right) \\ \log\left(\frac{P(Y_{i1} = 0, Y_{i2} = 1)}{P(Y_{i1} = 0, Y_{i2} = 0)}\right) = \alpha_{31} + \alpha_{32}\log\left(\frac{\widehat{P_{11}}}{\widehat{P_{00}}}\right) + \alpha_{33}\log\left(\frac{\widehat{P_{10}}}{\widehat{P_{00}}}\right) + \alpha_{34}\log\left(\frac{\widehat{P_{01}}}{\widehat{P_{00}}}\right) \end{cases}$$

where $\widehat{P_{ab}}$ is the estimated risk of $P(Y_{i1} = a, Y_{i2} = b)$ based on the prediction model under evaluation. The parameters $\alpha_{11}$, $\alpha_{21}$ and $\alpha_{31}$ represent the calibration-in-the-large and are estimated by fitting the above model with the slopes $\alpha_{12}$, $\alpha_{23}$ and $\alpha_{34}$ fixed at unity and the slopes $\alpha_{13}$, $\alpha_{14}$, $\alpha_{22}$, $\alpha_{24}$, $\alpha_{32}$, and $\alpha_{33}$ fixed at zero. The calibration slope for each joint outcome are $\alpha_{12}$, $\alpha_{23}$ and $\alpha_{34}$, which are estimated by fitting the above model with $\alpha_{13}$, $\alpha_{14}$, $\alpha_{22}$, $\alpha_{24}$, $\alpha_{32}$, and $\alpha_{33}$ fixed at zero.

# 11 Supplementary Figure

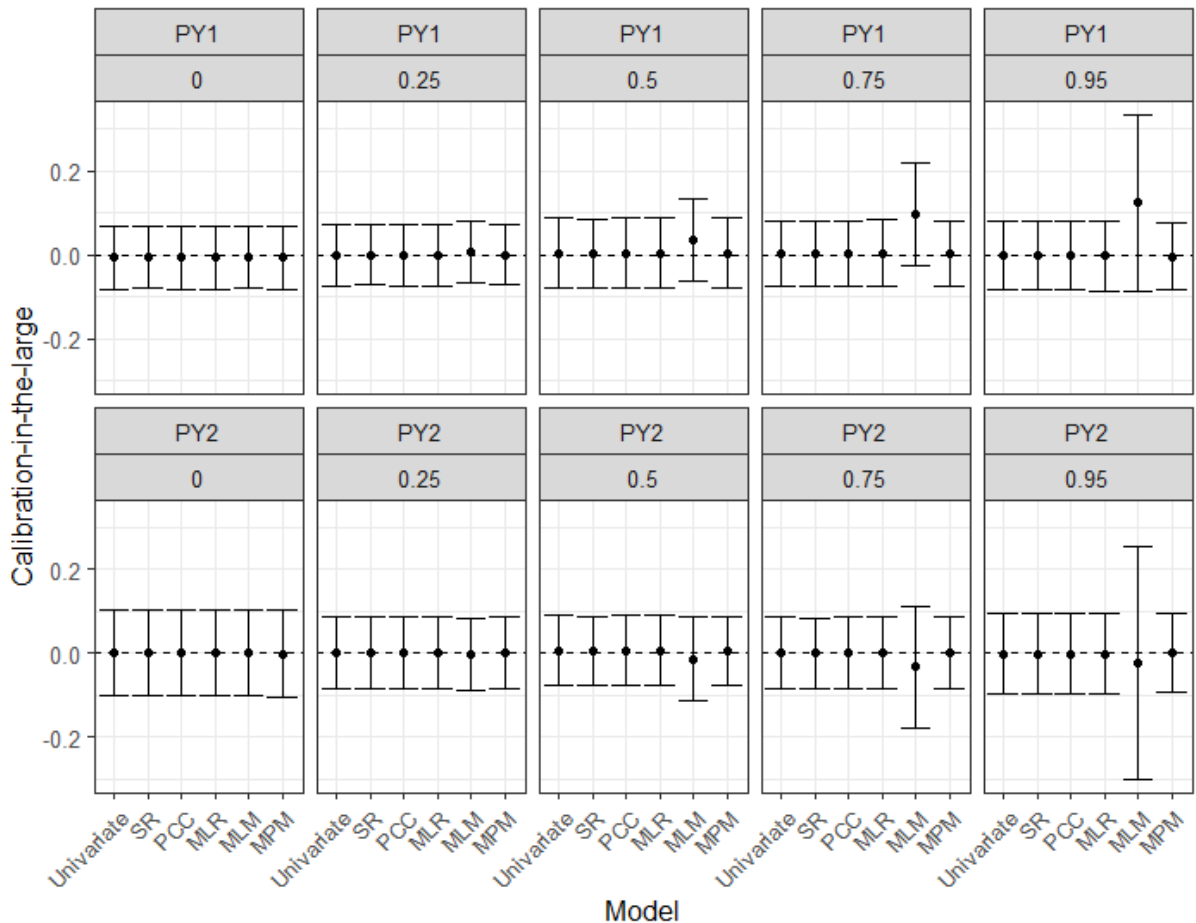

**Supplementary Figure 1:** Calibration-in-the-large for each model across all simulation scenarios. Each column of plots corresponds to a value of ρ, while each row of plots is a marginal outcome as follows: PY1 denotes $P(Y_{1i} = 1)$, and PY2 denotes $P(Y_{i2} = 1)$. The dashed horizontal lines show the reference value for the calibration-in-the-large of 0. Univariate = two independent CPMs, SR = stacked regression, PCC = probabilistic classification chains, MLR = multinomial logistic regression, MLM = multivariate logistic model, and MPM = multivariate Bayesian probit model.

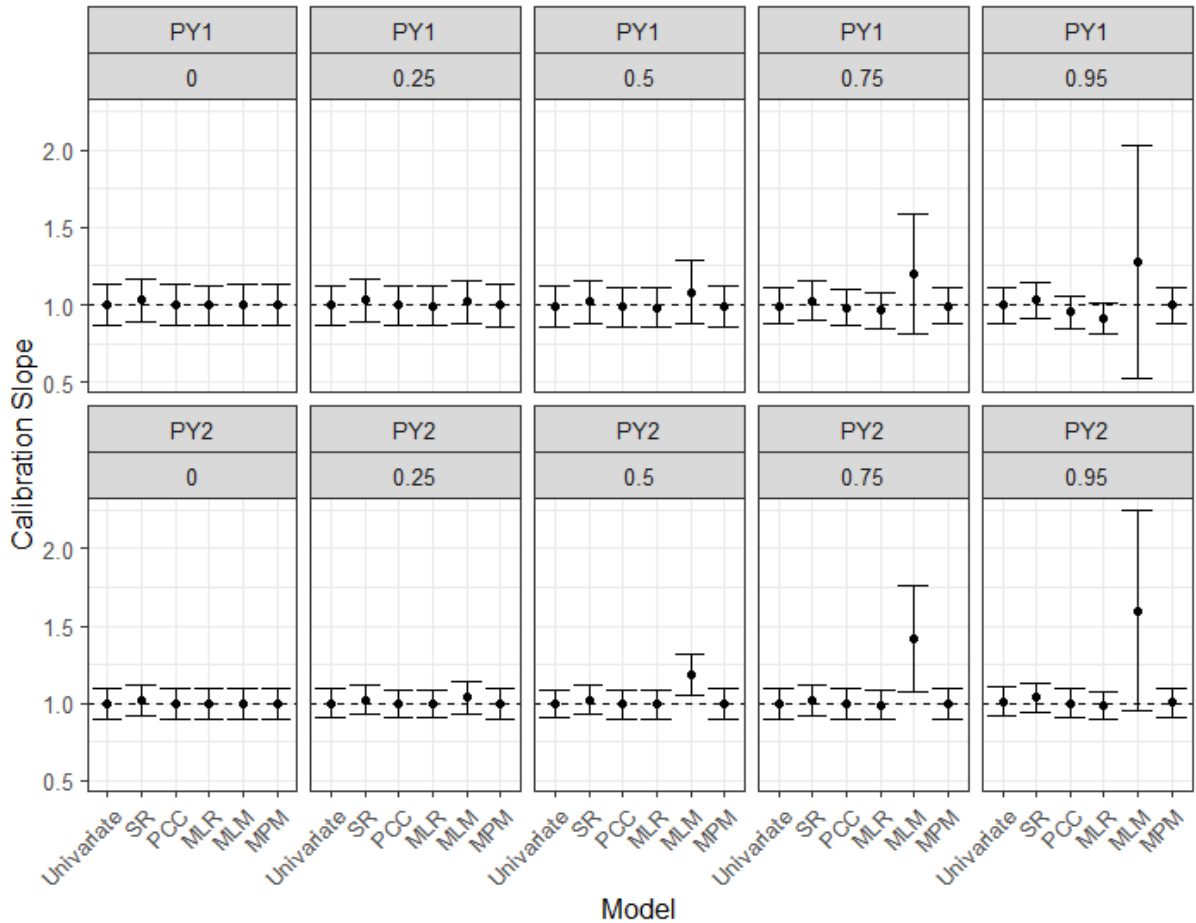

**Supplementary Figure 2:** Calibration slope for each model across all simulation scenarios. Each column of plots corresponds to a value of ρ, while each row of plots is a marginal outcome as follows: PY1 denotes $P(Y_{1i} = 1)$, and PY2 denotes $P(Y_{i2} = 1)$. The dashed horizontal lines show the reference value for the calibration slope of 1. Univariate = two independent CPMs, SR = stacked regression, PCC = probabilistic classification chains, MLR = multinomial logistic regression, MLM = multivariate logistic model, and MPM = multivariate Bayesian probit model.

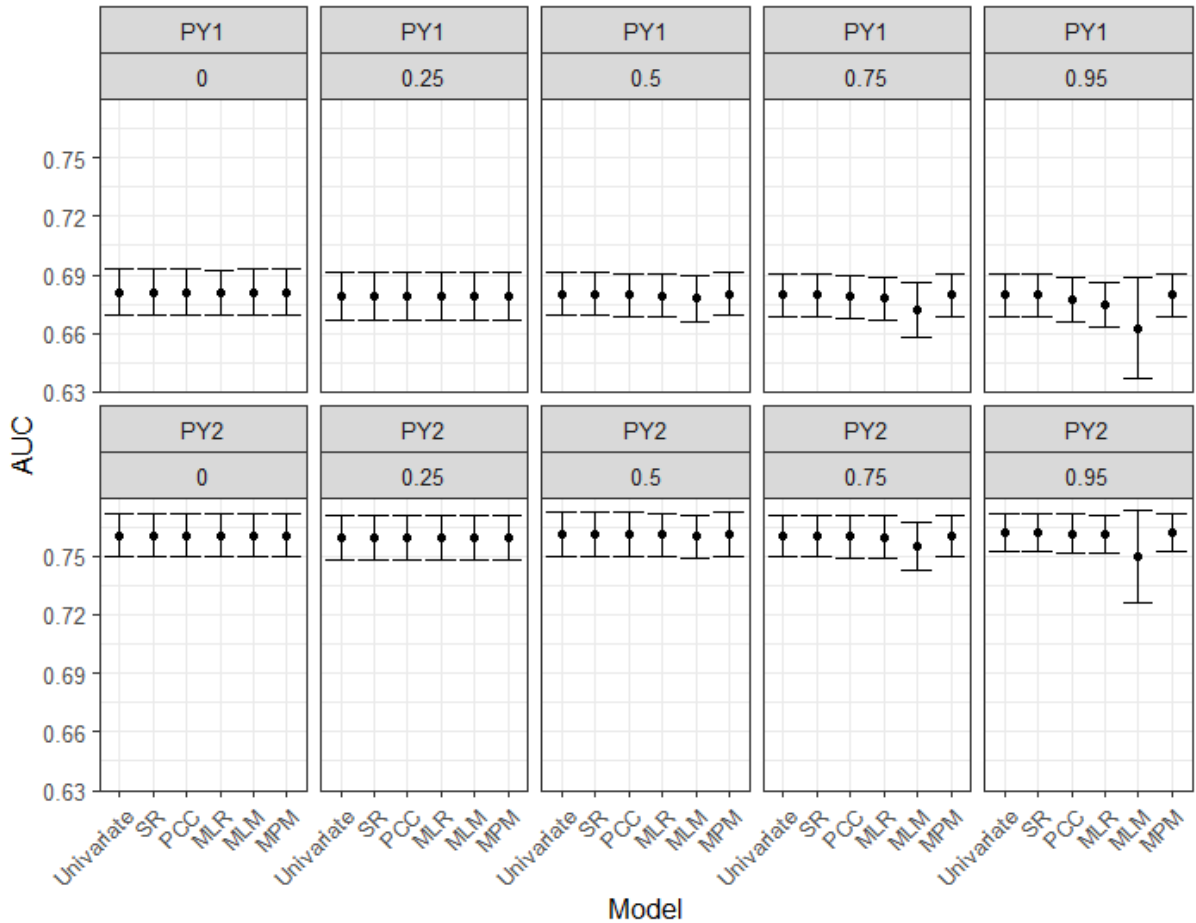

**Supplementary Figure 3:** Area under receiver operating characteristic curve (AUC) for each model across all simulation scenarios. Each column of plots corresponds to a value of ρ, while each row of plots is a marginal outcome as follows: PY1 denotes $P(Y_{1i} = 1)$, and PY2 denotes $P(Y_{i2} = 1)$. Univariate = two independent CPMs, SR = stacked regression, PCC = probabilistic classification chains, MLR = multinomial logistic regression, MLM = multivariate logistic model, and MPM = multivariate Bayesian probit model.

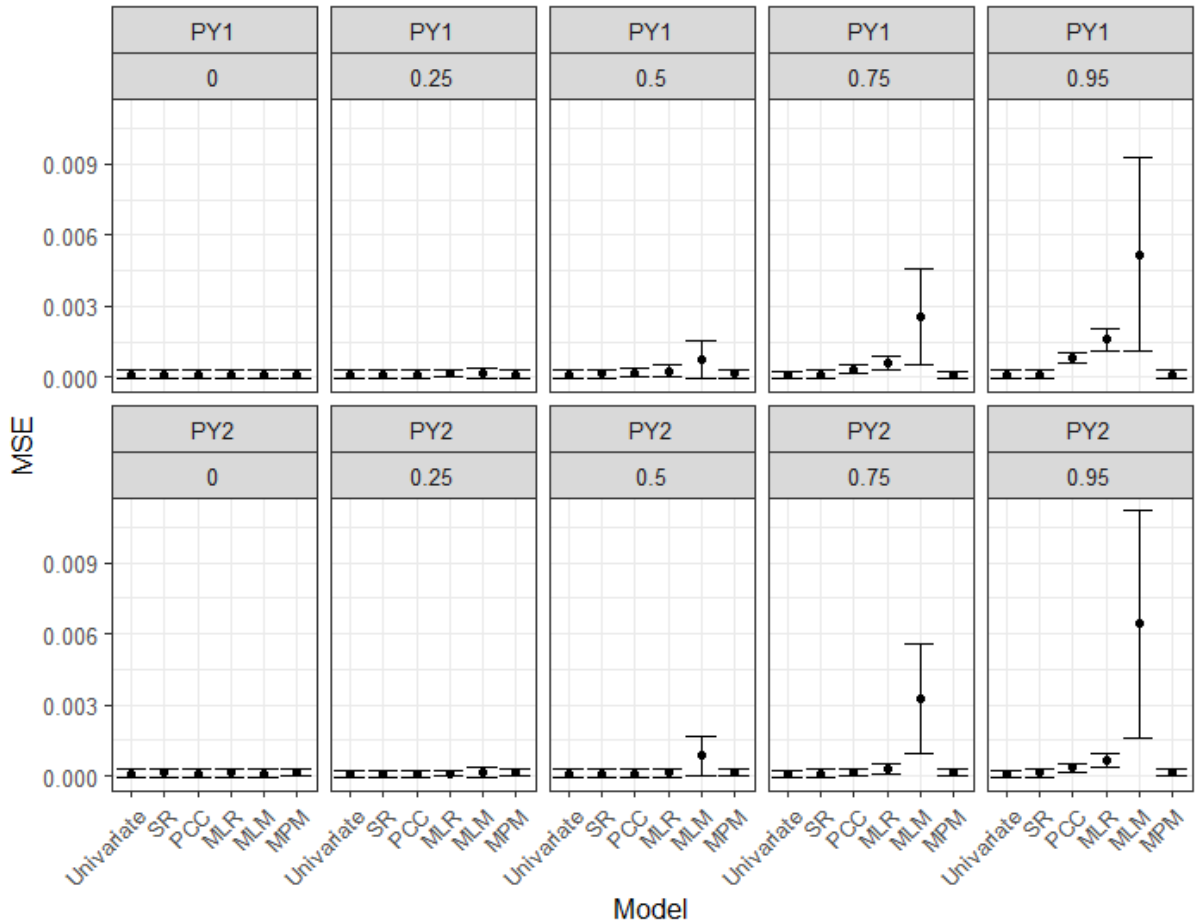

**Supplementary Figure 4:** Mean Squared Error (MSE) for each model across all simulation scenarios. Each column of plots corresponds to a value of ρ, while each row of plots is a marginal outcome as follows: PY1 denotes $P(Y_{1i} = 1)$, and PY2 denotes $P(Y_{i2} = 1)$. Univariate = two independent CPMs, SR = stacked regression, PCC = probabilistic classification chains, MLR = multinomial logistic regression, MLM = multivariate logistic model, and MPM = multivariate Bayesian probit model.

# 12  Supplementary Tables

**Supplementary Table 1**: The list of variables included in each analytical method for the empirical study of the MIMIC-III data

| **Variables** |
|---|
| Age |
| Gender |
| Indication of previous hospitalisation for a given patient, prior to the current ICU admission |
| Admission type (emergency vs. non-emergency) |
| Ethnicity (white vs. other) |
| Mean bicarbonate over first 24 hours of the ICU admission |
| Mean creatinine over first 24 hours of the ICU admission |
| Mean chloride over first 24 hours of the ICU admission |
| Mean haemoglobin over first 24 hours of the ICU admission |
| Mean platelet count over first 24 hours of the ICU admission |
| Mean potassium over first 24 hours of the ICU admission |
| Mean partial thromboplastin time over first 24 hours of the ICU admission |
| Mean international normalised ratio over first 24 hours of the ICU admission |
| Mean prothrombin time over first 24 hours of the ICU admission |
| Mean white blood cell count over first 24 hours of the ICU admission |
| Mean heart rate over first 24 hours of the ICU admission |
| Mean systolic blood pressure over first 24 hours of the ICU admission |
| Mean diastolic blood pressure over first 24 hours of the ICU admission |
| Mean respiration rate over first 24 hours of the ICU admission |
| Mean temperature (Celsius) over first 24 hours of the ICU admission |
| Mean Oxygen saturation (Sp02) over first 24 hours of the ICU admission |
| Mean glucose over first 24 hours of the ICU admission |